\documentclass{IEEEoj}
\usepackage{cite}
\usepackage{url}
\usepackage{amsmath,amssymb,amsfonts}
\usepackage{algorithmic}
\usepackage{graphicx,color}
\usepackage{textcomp}
\def\BibTeX{{\rm B\kern-.05em{\sc i\kern-.025em b}\kern-.08em
    T\kern-.1667em\lower.7ex\hbox{E}\kern-.125emX}}
\def\OJlogo{\vspace{-10pt}\includegraphics[height=20pt]{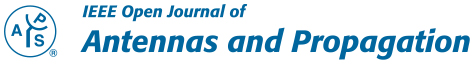}}
\usepackage{multirow}
\usepackage[table,xcdraw]{xcolor}
\usepackage{blindtext}
\usepackage{hyperref}

\begin{document}
\receiveddate{XX Month, XXXX}
\reviseddate{XX Month, XXXX}
\accepteddate{XX Month, XXXX}
\publisheddate{XX Month, XXXX}
\currentdate{XX Month, XXXX}
\doiinfo{OJAP.2020.1234567}

\title{An Update on the External Calibrator for Hydrogen Observatories (ECHO)}
%
%
%

\author{YIFAN ZHAO\IEEEauthorrefmark{1}, DANIEL C. JACOBS\IEEEauthorrefmark{1}, TITU SAMSON\IEEEauthorrefmark{1}, MRUDULA GOPAL KRISHNA\IEEEauthorrefmark{1}, MICHAEL HORN\IEEEauthorrefmark{1}, MARC-OLIVIER R. LALONDE\IEEEauthorrefmark{1}, RAVEN BRAITHWAITE\IEEEauthorrefmark{1}, AND LOGAN SKABELUND\IEEEauthorrefmark{1}}
\affil{Arizona State University, Tempe, AZ 85281 USA}
\corresp{CORRESPONDING AUTHOR: Yifan Zhao (e-mail: amyzhao@asu.edu).}
\markboth{An Update on the External Calibrator for Hydrogen Observatories (ECHO)}{Zhao \textit{et al.}}



\begin{abstract}
Precision measurements of the beam pattern response are needed to predict the response of a radio telescope. Mapping the beam of a low frequency radio array presents a unique challenge and science cases such as the observation of the 21\,cm line at high redshift have demanding requirements. Drone-based systems offer the unique potential for a measurement which is entirely under experimenter control, but progress has been paced by practical implementation challenges. Previously, a prototype drone system, called the External Calibrator for Hydrogen Observatories (ECHO), demonstrated good performance in making a complete hemispherical beam measurement. This paper reports updates to the system focusing on performance of a new drone platform, minimizing interference from the drone, and a new transmitter.
\end{abstract}


%

\maketitle

\section{Introduction}

\IEEEPARstart{M}{easurements} of the 21 cm sky background signal from Cosmic Dawn, the Dark Ages, and the Epoch of Reionization (EoR) rely on highly precise measurements of the sky with radio arrays. This background spectral line signal of the EoR is fainter, about 1000 to 10,000 times smaller, than the smooth spectrum foreground. These two are in principle distinguishable within the limitations imposed by uncertainty in the instrument response. The precision of this model is limited by, among others, knowledge of the antenna beam. 
Beam measurement techniques are an ongoing challenge in radio astronomy at long wavelengths, since antennas too large to be fully characterized in a laboratory setting. 

Typical beam calibration techniques present complications at long wavelengths. Anechoic chambers, used to calibrate high frequency antennas, are too small to fully enclose the antennas. Outdoor antenna ranges allow for measurements of lower frequency antennas, but do not capture environmental factors like soil variation or plants which happen in the as-built telescope. In-situ beam calibration is needed. Astronomical sources can be used if enough baselines are available to resolve them \cite{Pober}, \cite{Cox}. But with few exceptions, sources are not bright enough to be detected at high signal to noise in the sidelobes of a single antenna under test. Other promising methods include using satellite constellations \cite{Neben}, \cite{Line}, \cite{Chokshi} and inference using input reflection \cite{Monsalve}. Drone-based methods have also been explored and we focus on this option here. 

The External Calibrator for Hydrogen Observatories (ECHO) project aimed to address these issues for 21-cm calibration with a drone system optimized to map wide-field antennas operating at wavelengths of 2-4m to produce beam maps suitable for validating 21cm performance criteria. A first trial of this concept demonstrated precision better than 1\% across a spherical pattern \cite{Jacobs}. 

The type of measurement demonstrated was a far field pattern map measured on a sphere with a drone mounted transmitter acting as a probe. The antenna was a 1/4 wave dipole tuned to 1.7m mounted on a 2 meter ground plane. Measured at a radius of 100m, the transmitter was judged to be beyond the far-field of the antenna.   This first test demonstrated repeatability and accuracy similar to or better than satellite-based measurements, it also helped clarify the challenges and practical design requirements of such a system include ease of use. 

The most difficult practical matters were consistency of flight dynamics and characterization of self-made RFI. This  follow-up to Jacobs et al. 2017 documents some of the practical progress made since that time. A detailed study was made of requirements resulting in a new custom-built drone platform which is described in Section 2. Section 3  characterizes the radio frequency interference (RFI) emitted by the drone, and how it was minimized. Section 4  describes the flight performance of the system in a series of test flights at the Owens Valley Radio Observatory's Long Wavelength Array (OVRO-LWA) and in Tempe, Arizona. Finally, section 5 will describe the RF payload, specifically the design and development of a noise source and a modulator as a method to subtract residual self-interference.

\section{Design of a Low Cost Platform}

Following the first demonstration with a prototype in ref. \cite{Jacobs}, a new platform was needed to make routine measurements at operating arrays. Issues affecting the measurement quality included the limited flight time and  low angular stability. The system also needed to be robust enough for regular field use by a student team to make routine measurements at remote sites. Flight time, repairability, and mounting are all strongly dependent on the drone technology used.

The design process started with a trade study of off-the-shelf drones that fit the requirements. Generally, off-the-shelf systems are lower risk but more costly and difficult to customize. We selected a system by following a requirements definition and prototype evaluation process, evaluating six potential options and acquiring three (the Steadidrone Vader, the Yuneec H520, and the Chiropter) for test. The Chiropter was a custom design developed in collaboration with the ASU DREAMS robotics lab. 

The requirements included: high positional and angular stability ($<$10cm stability along the x, y, and z axes and a $<1^o$ angular stability) due to calibration sensitivity; a vibrationally robust system able to withstand the vibrational frequency of the motors; a 30-minute minimum flight time for beam mapping efficiency; a slow flight speed between 0.5 and 1 m/s for beam map resolution; and a 100m flight ceiling for far-field beam mapping of a single antenna, or near-field beam mapping of an array. In addition, to minimize gain changes as a function of path length difference and to minimize flight time, the drone must be able to follow a semi-spherical path, which must be precise enough to allow for a beam map with a resolution of 1 degree. As a result, the drone also needs to have a programmable flight path and time-tagged GPS information. Finally, the drone must be able to carry and mount a 600g payload at least 20cm in height, 30cm long, and 15cm wide. The payload mount must be flexible enough to allow for reasonable changes to payload components, which may include a radio antenna, a transmitter, and a chopper board. Room for additional inline RF components like filters and splitters should also be available. Table \ref{design_requirements} lists these requirements.

We initially expected the Vader, with its large frame and high performance propulsion system, to have the best flight time and mounting performance. However, its un-loaded hover time of 20 minutes was significantly below the Chiropter at 45 minutes. The interface program's unintuitive user interface and the inability to attach a custom payload to the Yuneec H520 also made it unsuitable. Ultimately, the low cost of Chiropter, which would allow us to manufacture several units, made it a clear choice for the ECHO system.

\begin{figure}[!t]
\centering
\includegraphics[width=3.5in]{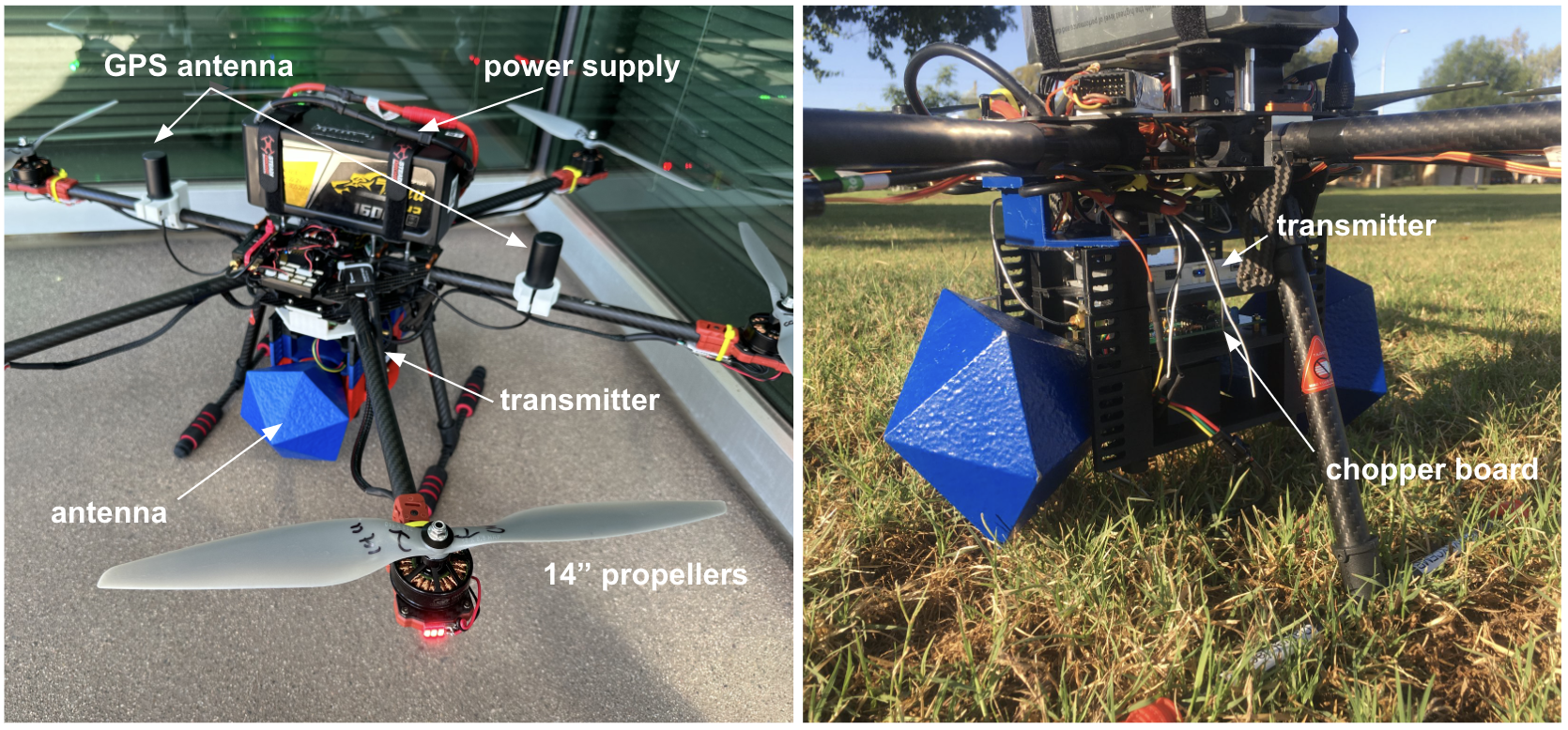}
\caption{The custom built ECHO drone, ready for flight. The Mauch power module sits just under the flight frame and the payload sits below that. A blue BicoLOG transmitting antenna sits at the bottom of the payload. For scale, the propellers are 14'' tip-to-tip. To the right a different angle shows the payload stacking tray system. Individual connectorized components are mounted in stackable trays which are bolted to a mounting plate beneath the drone with nylon all-thread. The antenna is mounted using a special tray with locking features that keep antenna from moving.}
\label{drone}
\end{figure}

The Chiropter uses a hexacopter configuration for stability and minimal weight, with six arms and six 14'' plastic propellers. The arms and body are constructed from carbon fiber, to be light enough ($<$1 kg) for 30 minutes of flight. The arms are a from a kit that has been modified to allow for larger propellers. For propulsion, the Chiropter uses six Cobra 4008-36 400Kv motors. The Chiropter uses the Pixhawk 2.1 flight controller and the Here+ RTK GPS set. The Frsky X8R Receiver, the Readymade X8R Cable, and Holybro 915MHz Radio are used for communications. For power, the Chiropter uses the 6S 16000mAh Battery and the Mauch 4-14s Battery Kit. Fig. \ref{drone} shows the Chiropter system, with the battery and custom payload mounted and ready for flight. The next sections discuss the battery and payload.

\begin{table*}[t]
\caption{ECHO Design Requirements}
\label{design_requirements}
\resizebox{\textwidth}{!}{%
\begin{tabular}{|
>{\columncolor[HTML]{EFEFEF}}l |l|l|l|l|l|}
\hline
{\color[HTML]{333333} \textbf{Drone}} & \cellcolor[HTML]{FFFFFF}\textbf{3D-Robotics X8} & \cellcolor[HTML]{FFFFFF}\textbf{Steadidrone Vader} & \cellcolor[HTML]{FFFFFF}\textbf{Yuneec H520} & \cellcolor[HTML]{FFFFFF}\textbf{Chiropter} & \cellcolor[HTML]{C0C0C0}\textbf{Relevance} \\ \hline
{\color[HTML]{333333} \textbf{Payload Capacity}} & 0.2 kg & 6.6 kg & 0.8 kg & \&gt;0.6 kg & \cellcolor[HTML]{FFCCC9} \\ \cline{1-5}
{\color[HTML]{333333} \textbf{Mounting}} & \begin{tabular}[c]{@{}l@{}}Possible, but little\\ space\end{tabular} & Easy and accessible & No & Easy and accessible & \cellcolor[HTML]{FFCCC9} \\ \cline{1-5}
\textbf{Scriptable Flight Path} & Yes & Yes & No & Yes & \cellcolor[HTML]{FFCCC9} \\ \cline{1-5}
\textbf{Time-Tagged GPS Data} & Yes & Yes & No & Yes & \cellcolor[HTML]{FFCCC9} \\ \cline{1-5}
\textbf{Availability} & No & No & Yes & Yes & \cellcolor[HTML]{FFCCC9} \\ \cline{1-5}
\textbf{Hover Time} & 15 min & 25 min & 25 min & 45 min & \multirow{-6}{*}{\cellcolor[HTML]{FFCCC9}Critical} \\ \hline
\textbf{Price} & \$1,300.00 & \$15,000.00 & \$4,000.00 & \$2,800.00 & \cellcolor[HTML]{FFCE93} \\ \cline{1-5}
\textbf{Repairable} & Possible but difficult & Very difficult & No & Easy & \cellcolor[HTML]{FFCE93} \\ \cline{1-5}
\textbf{Faults} & 3/40 & 1/1 & 0/2 & 1/25 & \multirow{-3}{*}{\cellcolor[HTML]{FFCE93}Important} \\ \hline
\textbf{Retractable Legs} & No & Yes & Yes & No & \cellcolor[HTML]{FFFC9E}Nice to have \\ \hline
\textbf{Weight (with battery)} & 5.5 kg & 15.0 kg & 1.7 kg & 3.9 kg & \cellcolor[HTML]{FFFFC7}Safety \\ \hline
\textbf{Year} & 2015 & 2017 & 2018 & 2019 &  \\ \hline
\end{tabular}%
}
\end{table*}

\section{Determining and Minimizing Platform Noise}
\begin{figure}[!t]
\centering
\includegraphics[width=0.4\textwidth]{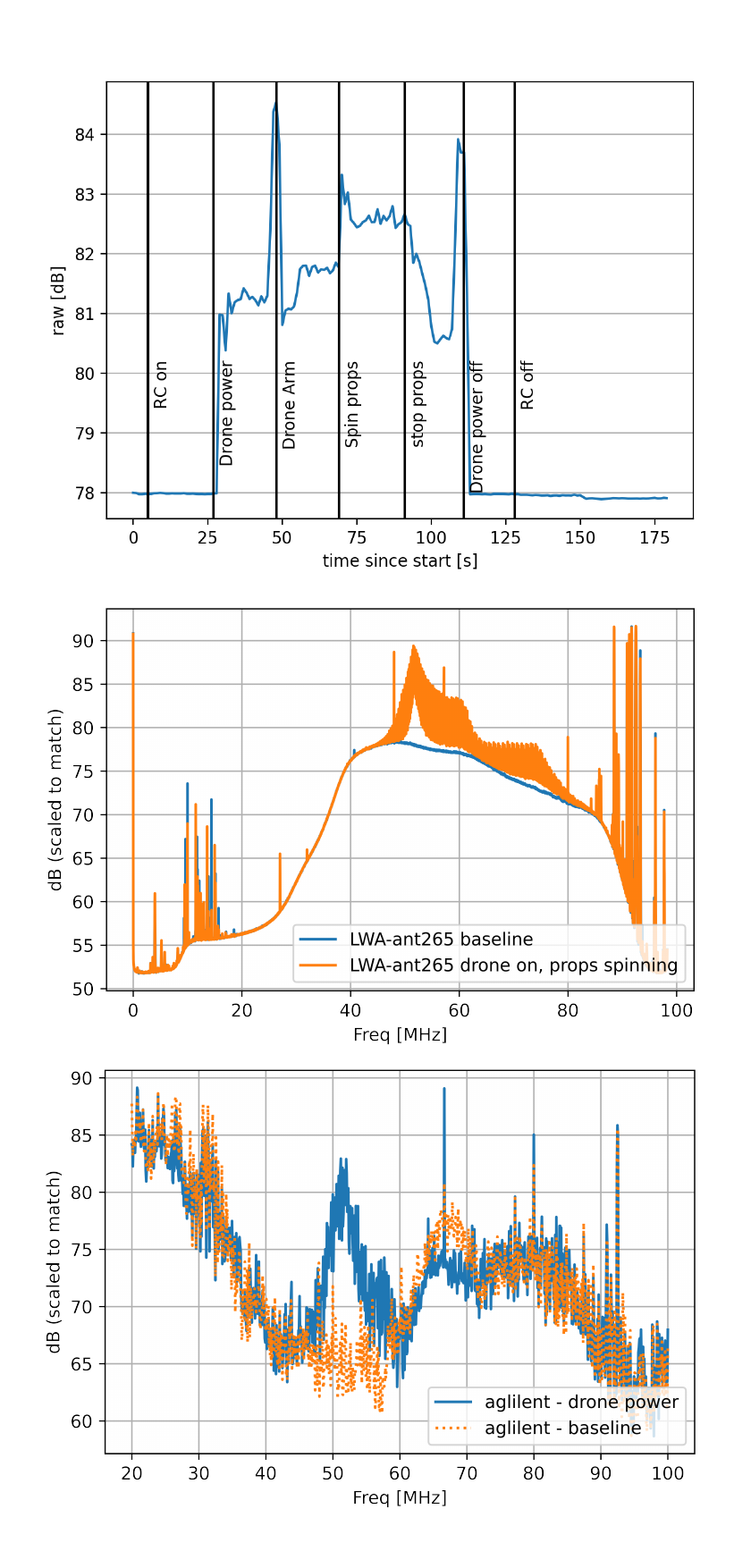}
\caption{(top) Sequential turn on test with drone on ground 15m from antenna 265 of the LWA, average of log power from 48 to 55MHz. (middle) Spectrum at 75s with the same field test setup as above. (bottom) Spectrum measured in the lab using a 4m wire antenna connected to a 6GHz Agilent spectrometer 10kHz RBW. Here the drone is operated on battery power and held within a few inches of the probe antenna. The motors are not running.}
\label{drone_rfi}
\end{figure}

Drone systems include switching power supplies, radios, and high current waveforms, all of which can generate RFI. During a field test of the ECHO system at the OVRO-LWA, RFI significantly higher than the baseline background noise was observed when the drone was turned on. Fig. \ref{drone_rfi} shows the RFI measured in the field and reproduced in the lab. The OVRO-LWA system observes between 13 and 90MHz and is sky noise limited across most of this band.  The spectrum recorded with the drone flying at approximately 100m altitude directly over the array shows several strong features across the band. These features do not appear in the spectrum taken soon after the drone was turned off. The primary feature is a comb with a prominent peak at 50MHz. Emission was confirmed with a measurement using a laboratory spectrometer, also in Figure \ref{drone_rfi}.  

The emission was further narrowed down by sequentially turning on drone systems while holding the drone next to the LWA antenna under test and measuring the observed power at 50MHz. This is the top field in Figure \ref{drone_rfi}. The three strongest changes occur when powering the flight computer and then arming and disarming. Surprisingly, spinning the propellers contributed a relatively small power increase. This drew attention away from the motors and to the other systems and ultimately to a switching power supply. 

The drone system has two power circuits: a high voltage system which runs directly to motors, and a 5 volt system which powers the flight computer and radios. The payload was powered by an independent battery in this test. The motors are driven by large FETs at speeds of order kHz. The 5V power supply is a buck-mode switching regulator that runs at around 300kHz. These power supplies are typically integrated into the drone high voltage battery distribution harness and also provide current and voltage telemetry. RFI testing of the X8 power module using near field probes narrowed down the source of the RFI to the MOSFET on the board.

Measurement of undesirable emission was found to be difficult to reproduce. Factors at play included small changes in probe device geometry, radio background in the lab, and even the geometry of nearby metal surfaces like tabletops. Not all fields measured at close distances translate into power radiated into the far field. However it is reasonable to suppose that components with weaker field strengths radiate less. A magnetic field probe was made by forming a loop of wire dipped in liquid insulator to form a rigid search coil, which was connected to a Fieldfox spectrum analyzer. A repeatable test setup was devised using the magnetic field probe connected to the spectrum analyzer, table clamps, and a dedicated tabletop. With the probe and the device under testing clamped securely, measured spectra were seen to be reliably repeatable. We found  field strengths to vary strongly with position of both the probe and even the position of the part in the room. These tests were prototyped in the lab and then repeated in an RFI shielded room.

RFI from the power circuit can be mitigated by shielding, replacing the power supply with a similar unit with lower EMI, or building a custom unit optimized to produce acceptable noise levels that are undetectable above the background. Three off-the-shelf replacement power supplies and three attenuation methods were tested, and multiple combinations reached acceptable levels. 

\begin{figure}[!t]
\centering
\includegraphics[width=0.5\textwidth]{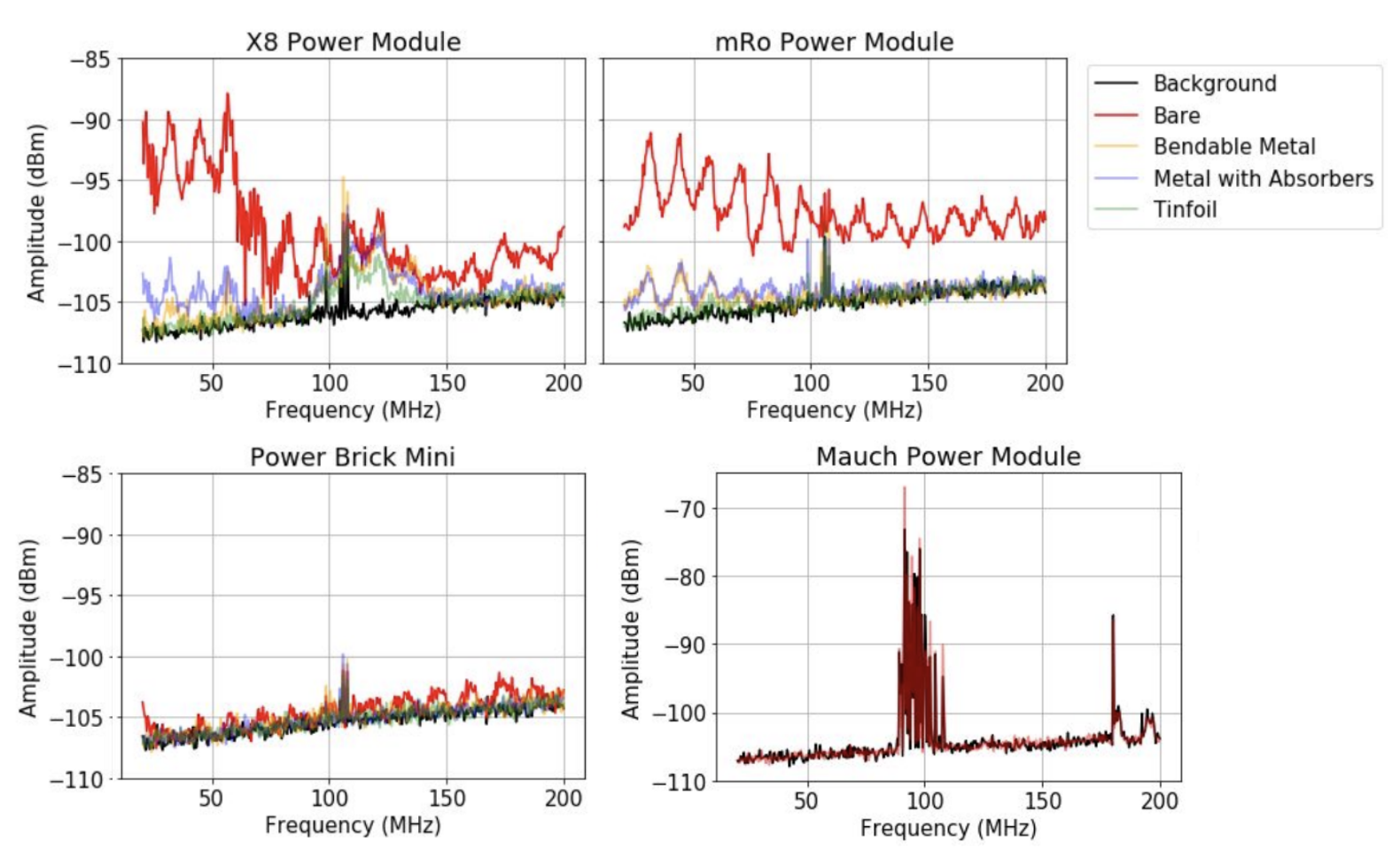}
\caption{Comparison of results of the different alternate power modules, each referenced again the background. Different shielding methods were also tested and shown here, though a simple tinfoil wrapping was agreed upon as the final implementation.}
\label{rfi_testing}
\end{figure}

Without any shielding, the mRo, X8, and Power Brick Mini power modules showed emission above the background, while the Mauch is the quietest. With the addition of a simple foil wrapping shielding technique, the EMI becomes indistinguishable from the background. Detailed results are shown in Fig. \ref{rfi_testing} and table \ref{rfi_results}, which lists the mean power above the background in dBm for each of the the alternate power modules. Given the results, the Mauch power module was chosen. Preliminary testing suggests that radiation from the power module is no longer the dominant concern, but more testing is needed to fully characterize the drone emissions. This is described in the next section.

\begin{table}[!t]
\renewcommand{\arraystretch}{1.3}
\caption{RFI testing of different power modules, with and without shielding}
\label{rfi_results}
\centering
\begin{tabular}{|l|llll|}
\hline
 & \textbf{X8} & \textbf{mRo} & \textbf{Mini} & \textbf{Maunch} \\
\hline
\textbf{Powered On} & 6.6 dBm & 7.4 dBm & 0.9 dBm & 0.0 dBm \\
\textbf{Tinfoil shielding} & 0.8 dBm & 0.3 dBm & 0.2 dBm & 0.2 dBm \\
\hline
\end{tabular}
\end{table}

\subsection{Use of Chopper to Measure and Remove Noise}
Drones can cause backgrounds which are difficult to distinguish from the transmitter. If the transmitter is sufficiently narrow band and the drone background noise is relatively stable over a broad band, one could use nearby channels to determine the background. But a compressed or a broadband signal would make using nearby channels impossible. A chopper board was developed that turns the RF signal on and off to aid in background characterization. 

The basic design was for a SPDT switch which could route power from the input to the antenna or to a dummy load. The signal could be either ON (antenna), OFF (load),  or switching at a constant period. A further design requirement was that the switch be controllable by the drone pilot, robust, low power, and with no high speed clocks. These requirements were met by a design using relatively monolithic logic devices.  Control is made via the drone controller receiver which supplies a PWM signal proportional to a switch on the pilots controller.  A solid state switch supplies the primary RF function. Though such switches have been commonly used in other applications, this was a first use for 21cm cosmology applications and it was being considered as a replacement technology for mechanical switches currently in use by global experiments for absolute calibration. 

To determine requirements for switch time of the chopper board, drone speed was assumed to be about 0.3 degrees/second at a range of 100m. Aiming for a beam map of resolution of about 1 degree and for 30 samples per pixel would amount to a sample rate of 10 Hz. Ideally, the chopper board would switch between the ``off" and ``on" states at the same sample rate, so both states are evenly sampled. This determines a switch time requirement of 100ms.

\begin{figure*}[!t]
\centering
\includegraphics[width=\textwidth]{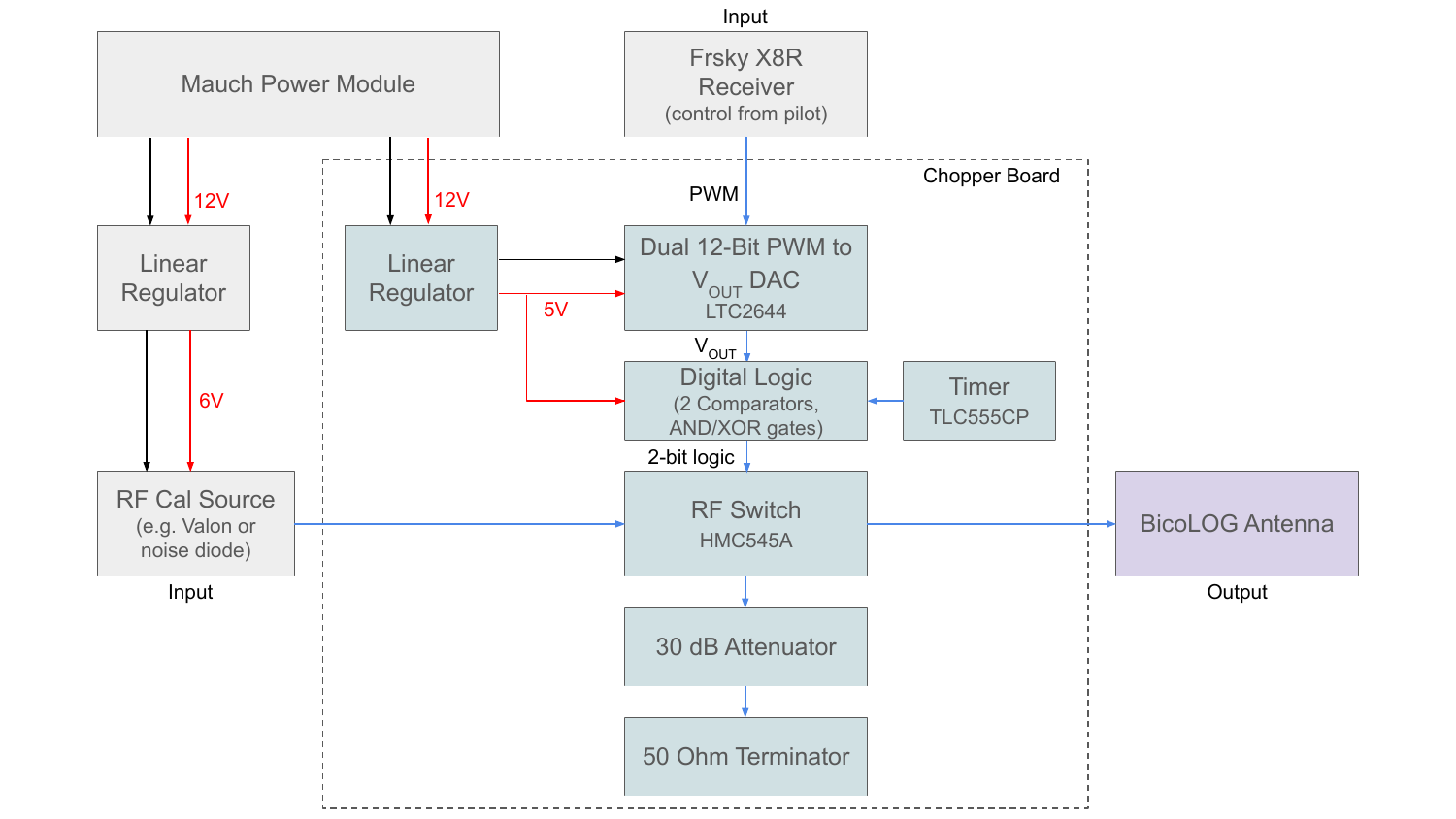}
\caption{Block diagram of RF Chopper, which switches the output of the board between the BicoLOG antenna and a terminated load. There are 3 main inputs to the Chopper: 1) the X8R Receiver, which sends PWM signals to the board based on controller switch position; 2) the Mauch Power Supply, which delivers 12V to the board; and 3) the Valon Transmitter, which sends our RF output signal into the board.}
\label{chopper_blockdiagram}
\end{figure*}

Fig. \ref{chopper_blockdiagram} shows the design of the chopper board, which switches the RF signal back and forth between the antenna and a dummy load. When routed to the dummy load, some signal will inevitably leak out. This level must be below the level of the drone background which we are trying to characterize. In initial bench testing using a Valon signal generator and a Fieldfox spectrum analyzer at the output of the chopper board, the ``off" state was determined to isolate the RF signal by at least 40dB, which could potentially be improved with better shielding around the board.

Flight testing of the chopper board revealed unintended behavior of the switch, where the ``on" state gain would change by as much as 5dB between two stable states. This suggested some instability in the connection of the electrical path. This instability was common during flight but difficult to reproduce on the bench. The problem was traced to the grounding of the HMC545A solid state RF switch. The switch only had one ground connection point. Initially, following a build rule to keep digital and RF grounds separate, the switch was grounded to the digital ground only. Connecting the RF ground to the digital ground eliminated the stability issue.

\section{Flight System Performance}
The Chiropter drone performed well in initial test flights. It was first deployed at the LWA Sevilleta in 2019. There, the only issue was a failure of the automated flight guidance which necessitated a manually controlled landing. This issue was traced to a redundant GPS causing an overcurrent fault leading to a reboot of the flight system, which surprisingly did not lead to a crash, only a disengagement of the autopilot. Flights at observatories were suspended during the pandemic and resumed in 2022 at Owens Valley Radio Observatory (OVRO). Three Chiropters were available for use. During test flights at OVRO, the flight system experienced several crashes, which was hypothesized to be due to motor failure. With telemetry indicating that motor control signals were saturated, motor failure was suspected.

\begin{figure}
\includegraphics[width=0.9\columnwidth]{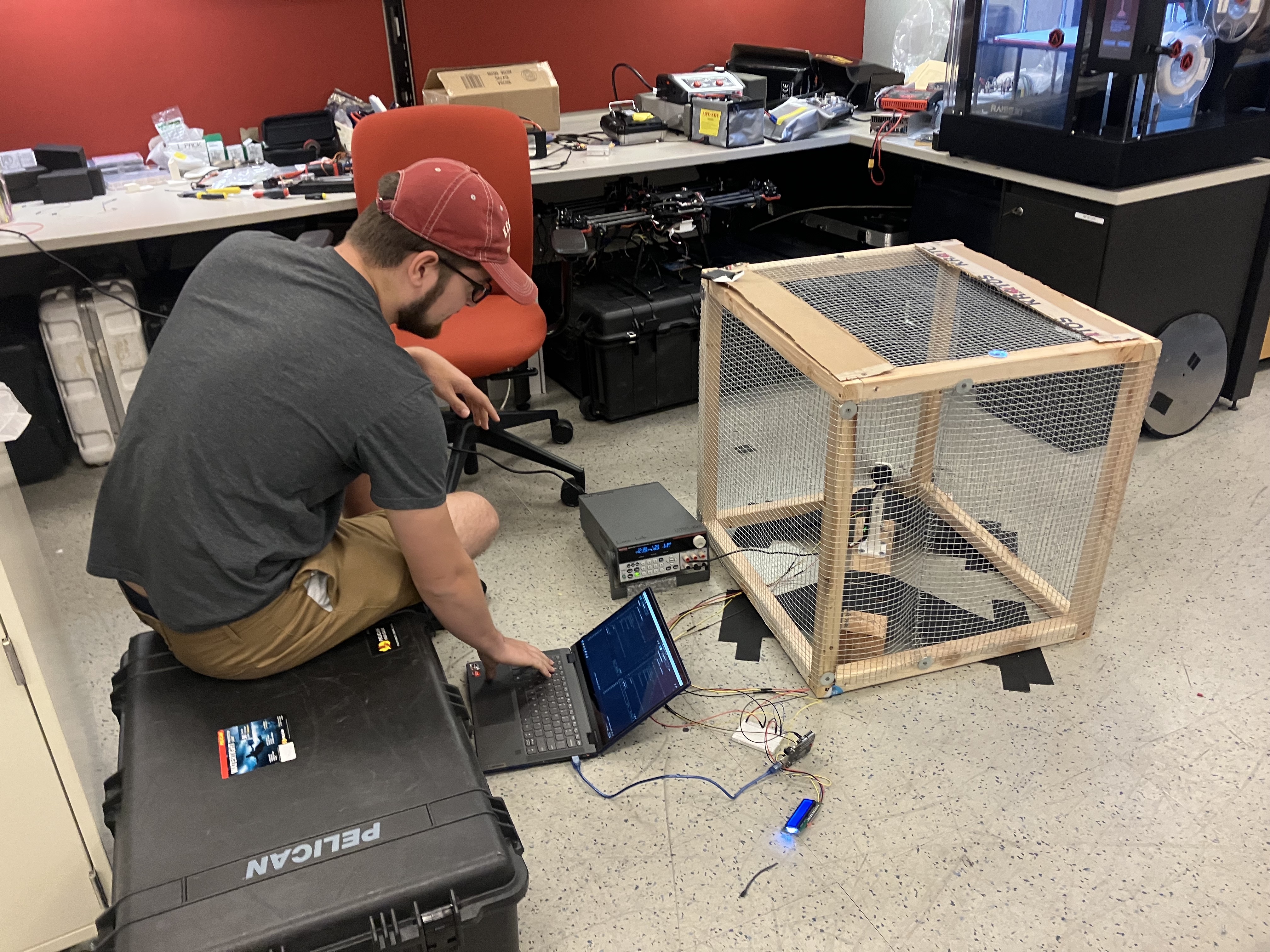}
\includegraphics[width=0.9\columnwidth]{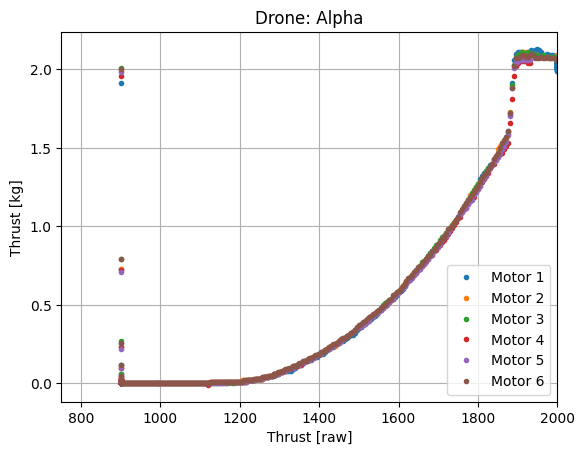}
\caption{Degredation of the propulsion system was investigated by testing motor, prop, and ESC combination on a thrust stand (top).  A calibrated load cell measured thrust while an arduino microcontroller drove the ESC through a range of commanded thrust levels. All motors on all drones were tested. Thrust was very consistent between motors. Thrust was also stable for a suitably long interval of 5 minutes. }
\label{fig:thrust_stand}
\end{figure}

Degradation of performance in the motors was investigated with an experimental thrust stand setup shown in Figure \ref{fig:thrust_stand}. A calibrated load cell measured thrust developed by a motor and prop combination while an arduino microcontroller drove the ESC through a range of commanded thrust levels. An example is shown in Figure \ref{fig:thrust_stand}. The entire apparatus was enclosed in stout wire mesh for safety. Motors and propellers displayed a steady output of thrust through the longest test lasting more than five minutes.

The Cobra 4008-36 400Kv motors running 14" props develop 1.2kg of thrust at full power and 800g at 50\%. The 6 propeller system generates 4.8kg of thrust at 50\% throttle. With a gross weight (including payload) of 2.2kg, the system is theoretically capable of 1.18g upward acceleration. This follows a commonly repeated rule of thumb which holds that thrust should be double the weight 50\% throttle.

\begin{figure}[!t]
\centering
\includegraphics[width=3in]{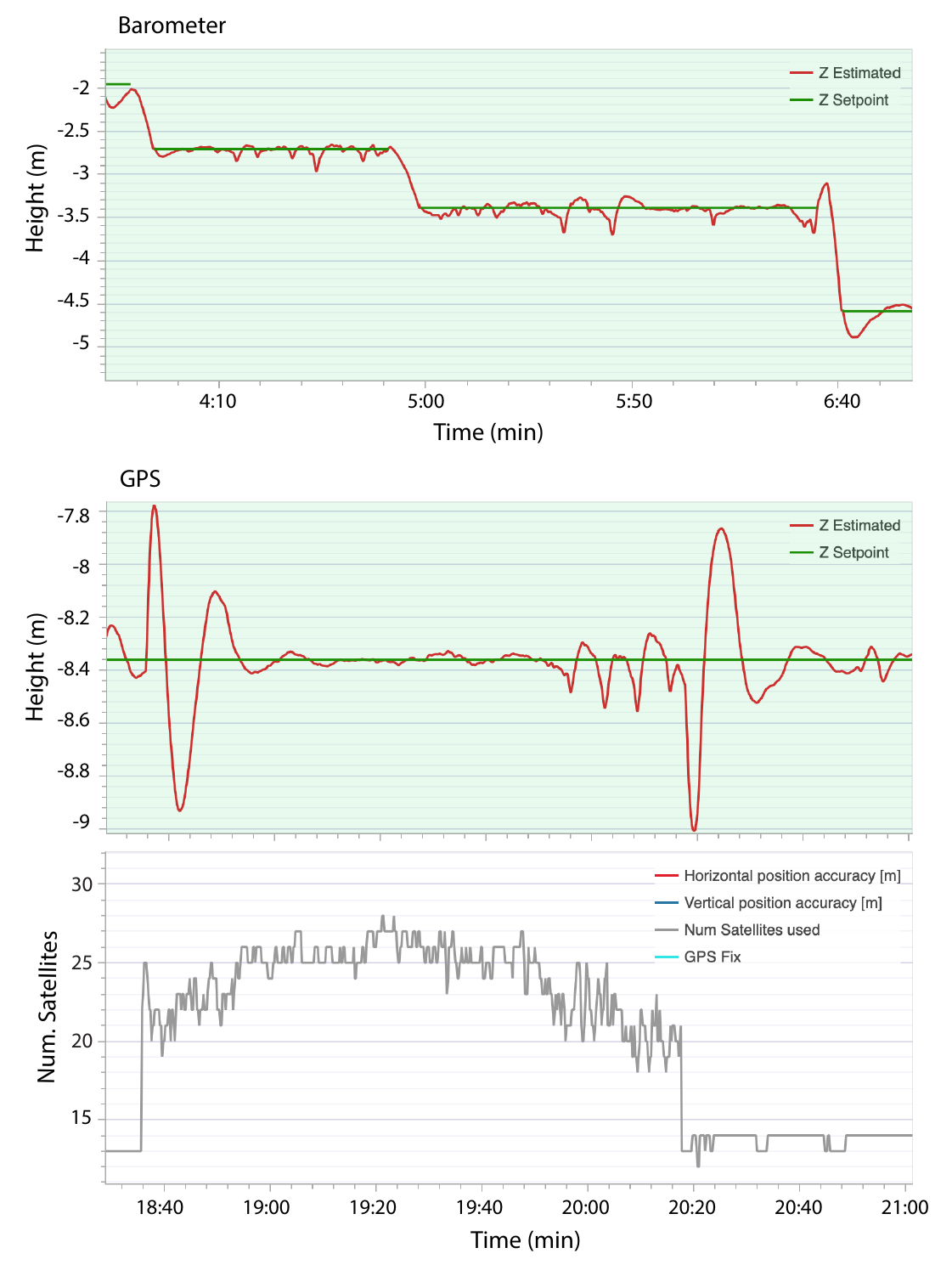}
\caption{(top) Commanded (in green) vs. actual (in red) altitude using barometer measurements as reference. (middle \& bottom) Commanded vs. actual altitude using GPS measurements as reference. Height is defined as negative in this flight controller. While using the GPS, uncommanded drops of up to 60cm are common while less than 30 using the barometer. Though there are always more than 15 GPS satellites and no loss of fix was warned, we do note that the largest height excursions while using GPS occur during times when fewer satellites are seen.}
\label{altitude}
\end{figure}

Other possible causes for the motor control issues could be either inaccurate accelerometer reading or inaccurate altitude reading (from the barometer or the GPS). Both these issues were investigated. Different foams for vibrational isolation of the accelerometer were tested and showed a significant improvement in the drone's altitude control. To further refine the altitude control of the drone, settings for altitude control on the flight controller were switched between barometer readings to GPS readings. A combination of these changes lead to steady test flights. Fig. \ref{altitude} shows the accuracy of the altitude control using barometer references vs. using GPS references. The accuracy of altitude control using a GPS reference is largely dependent on the number of satellites that are used for the real time kinetic (RTK) GPS system. Future work will focus on refining the use of GPS and improving vibrational isolation.

\section{RF Payload}
\subsection{Payload Mechanical Configuration}
Several issues identified during prototype testing could be solved by improvements in the payload mechanical design. In prototyping, the antenna was suspended by mono-filament fishing line in hopes that putting the antenna further from the drone would minimize any distortion of the antenna beam. It was also hoped that in the event of a crash, the line would break before the antenna. In practice the antenna wobbled independently of the drone meaning the measured angle of the drone could not be used to remove variation due to rotation of the transmitter beam. Another practical problem that emerged was the need to quickly and efficiently set up different payload variations including different numbers of components and easily attachable to new drone chassis. The mechanical solution also needed to be non conductive and lightweight.


The solution was a stackable set of trays clamped by nylon all-thread to an adaptor base which attaches to the bottom of the drone.  The adaptor mount points were deliberately kept weak so as to be the first failure point in the event of a crash. This causes the entire payload stack to separate from the drone and minimizes torques or other forces. This has been tested several times in real scenarios.  The tray supports were given slotted extrusions for fixing lashing cables or other inline RF components. The trays were 3D printed with SLA and weigh between 40 and 60 grams depending on height. The fully stacked payload is shown in Fig. \ref{drone}. The mechanical design has been iterated a dozen times with improvements to make it lighter and more flexible.  Design files are open source and available on the \href{https://github.com/dannyjacobs/ECHO/hardware}{ECHO Github}.

\subsection{Broadband Transmitter}
Though a narrow band tone generator was used as a convenient and robust calibration signal, it does not meet all requirements. Namely the requirement to measure the beam across a useful bandwidth, which for a 21cm experiment is at least 20~MHz.  A new noise source was developed with the goal of making measuring 60 to 80MHz simultaneously.  

\begin{figure}[!t]
\centering
\includegraphics[width=3in]{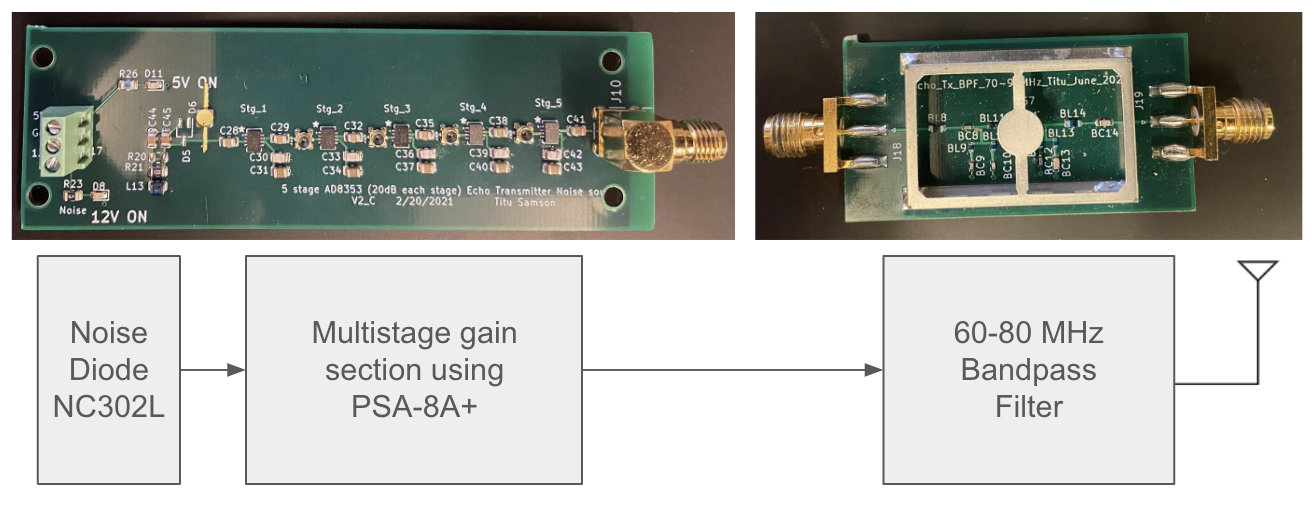}
\caption{Block diagram of the RF payload noise transmitter.}
\label{noise_block}
\end{figure}

\begin{figure}[!t]
\centering
\includegraphics[width=2.5in]{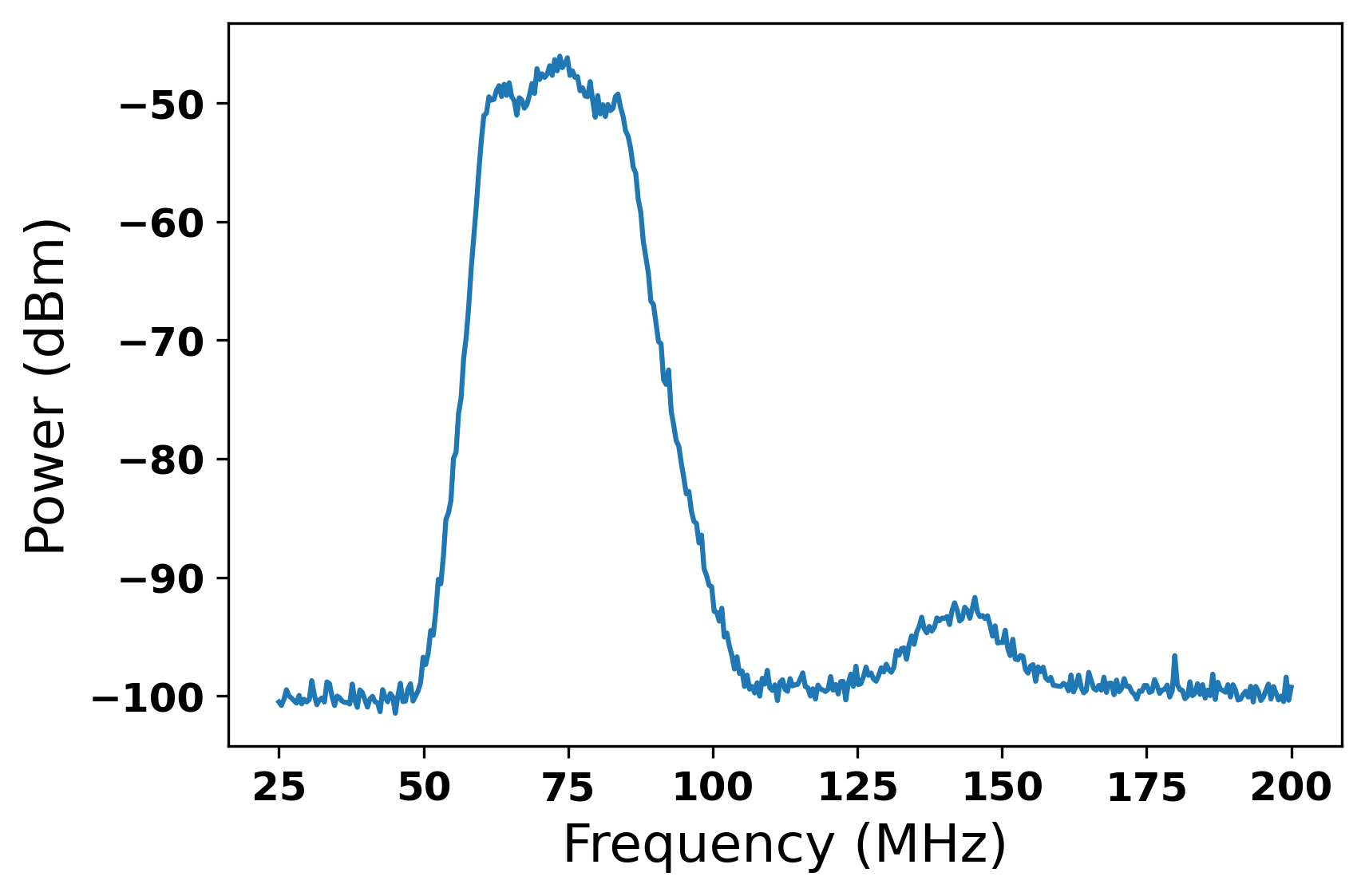}
\caption{Power spectrum of the noise transmitter, at the output of the bandpass filter. The secondary bump at 125-150 MHz is due to harmonics from the bandpass filter. Data was taken with a FieldFox spectrum analyzer set to a channel resolution of 100 Hz.}
\label{noise_trans}
\end{figure}

The resulting transmitter uses a noise diode cascaded with a multi-stage gain block and a bandpass filter. The noise diode was chosen for its ability to generate white noise across the desired frequency range, and the multi-stage amplifier was designed to provide XXX dBm in a 20kHz channel to achieve a power level similar to the Valon system. A bandpass filter with a center frequency of 70MHz and a bandwidth of 20MHz suppresses out-of-band noise. The noise diode was connected to input of the amplifier stages, and the bandpass filter was integrated at the output, as illustrated in Fig. \ref{noise_block}, giving a power spectrum output shown in Fig. \ref{noise_trans}.

The transmitter outputs -80 dBm/Hz or about -30dB per 100kHz channel. For comparison the Valon transmitter's monochromatic 0dB tone summed over the same 100kHz channel provides about -55dBm.  Subsequent gains or losses amount to roughly -2 dB of cable loss, 1 dB of antenna gain, and -50 dB of path loss for a estimated incident radiated power at the AUT of about -129dB/Hz and a total power of -55dBm, which is below the input 1dB compression point of an LWA antenna of -18 dBm and above the noise temperature of 300 K \cite{Ellinson}.

\section{Further Work}
Further work on the ECHO system will focus on refining the system for durability and ease of use, for regular field testing. Flight testing will aim to develop repeatable stability for height control, by using updated vibrational isolation for our accelerometer, refined usage of the GPS altitude reference, and testing of different control algorithms. Testing will continue at the OVRO-LWA, with the aim of a beam map of the farfield beam response of a single LWA antenna.

The far field distance depends on the physical size of the antenna that is being measured. Height constraints on the drone may or may not put the drone in the farfield. The far field distance depends on the physical size of the antenna which could include the distant surroundings depending on the level of precision needed by the science case. Current development has focused on amplitude-only measurements and it is assumed that the measured pattern reflects the far field. However any large scale effects could violate this assumption and require a near-field approach which would require additional measurements such as phase or a phase-less methods which require more elaborate spatial sampling.

Notes on the build of the ECHO system are available in the ECHO Memo Series at  \href{https://dannyjacobs.github.io/ECHO/memos/}{ECHO Memo Series} and the \href{https://github.com/dannyjacobs/ECHO}{ECHO Github}.

\section*{Acknowledgments}
Thanks to the International Drone Beam Interest Group (Chris Dupuis, Eamon Egan, Larry Herman, Cynthia Chiang, Will Tyndall, Chris Barbarie and others) for taking our systems to new latitudes and making everything work so much better.  Emily Kuhn, Akshatha Vydula, and Katherine Elder cheerfully slaved away at Owens Valley which would not have been possible without generous help from Mark Hodges and the OVRO staff. Thanks to Tom Bombadil for tree analysis which was not used but appreciated nonetheless. Testing at LWA-Sevellita was generously supported by Chris Dupuis, Jayce Dowell, and Greg Taylor.   This work was supported by the National Science Foundation under AST-2144995.

\ifCLASSOPTIONcaptionsoff
  \newpage
\fi



%

\end{document}